\documentclass[twoside]{dis07}
\usepackage[latin1]{inputenc}
\usepackage[dvips]{graphicx,epsfig,color}
\usepackage{wrapfig,rotating}
\usepackage{amssymb,amsmath,array}
\usepackage{cite}
\usepackage{./mcite}

\chardef\usc=95
\chardef\til=126
\catcode`\@=11 % @ signs are now treated as letters
\DeclareRobustCommand\xdotspace{\futurelet\@let@token\@xdotspace}
\def\@xdotspace{%
  \ifx\@let@token.\else
  \ifx\@let@token\bgroup.\else
  \ifx\@let@token\egroup.\else
  \ifx\@let@token\/.\else
  \ifx\@let@token\ .\else
  \ifx\@let@token~.\else
  \ifx\@let@token!.\else
  \ifx\@let@token,.\else
  \ifx\@let@token:.\else
  \ifx\@let@token;.\else
  \ifx\@let@token?.\else
  \ifx\@let@token/.\else
  \ifx\@let@token'.\else
  \ifx\@let@token).\else
  \ifx\@let@token-.\else
  \ifx\@let@token\@xobeysp.\else
  \ifx\@let@token\space.\else
  \ifx\@let@token\@sptoken.\else
   .\space
   \fi\fi\fi\fi\fi\fi\fi\fi\fi\fi\fi\fi\fi\fi\fi\fi\fi\fi}
\catcode`\@=12 % @ signs are no longer letters

\newcommand{\stru}[2]{%
   \relax\ifmmode\hbox{\vrule height#1 depth#2 width0pt}%
   \else\vrule height#1 depth#2 width0pt\fi}
\newcommand{\Ronum}[1]{\uppercase\expandafter{\romannumeral#1}}
\newcommand{\ronum}[1]{\expandafter{\romannumeral#1}}
\DeclareRobustCommand{\LaTeXZ}{%
  \LaTeX\kern-.05em4\kern-.1em
  {\raisebox{-0.2ex}{$\scriptstyle\text{ZEUS}$}}\xspace}
\DeclareMathAlphabet{\mathbf}{OT1}{cmr}{bx}{sl}
\newcommand{\eVdist}{\kern-0.06667em}

\newcommand{\gev}{{\,\text{Ge}\eVdist\text{V\/}}}

\newcommand{\pb}{\,\text{pb}}

\newcommand{\slashfrac}[2]{%
  \raisebox{0.5ex}{\ensuremath #1}\kern-0.12em/\kern-0.08em
  \raisebox{-.8ex}{\ensuremath #2}}
\newcommand{\sqr}[3]{%
    {\vcenter{\hrule height.#3ex\hbox{\vrule width.#2ex height#1ex
     \kern#1ex\vrule width.#3ex}\hrule height.#2ex}}}

\catcode`\@=11 % @ signs are now treated as letters
\newcommand{\parenbar}{\mathpalette\p@renb@r}
\def\p@renb@r#1#2{\vbox{%
  \ifx#1\scriptscriptstyle \dimen@.7em\dimen@ii.2em\else
  \ifx#1\scriptstyle \dimen@.8em\dimen@ii.25em\else
  \dimen@1em\dimen@ii.4em\fi\fi \offinterlineskip
  \ialign{\hfill##\hfill\cr
    \vbox{\hrule width\dimen@ii}\cr
    \noalign{\vskip-.3ex}%
    \hbox to\dimen@{$\mathchar300\hfil\mathchar301$}\cr
    \noalign{\vskip-.3ex}%
    $#1#2$\cr}}}
\catcode`\@=12 % @ signs are no longer letters

\newcommand{\IP}{{\rm I$\kern-0.01667em$P}\xspace}

\mathchardef\qsm=63
\mathchardef\pls=43
\mathchardef\mns=512
\mathchardef\plm=518
\mathchardef\eql=61
\mathchardef\smallleft=300
\mathchardef\smallright=301
\mathchardef\les=316
\mathchardef\gre=318
\mathchardef\leq=532
\mathchardef\grq=533
\catcode`\@=11 % @ signs are now treated as letters
\newcounter{pict@width}
\newcounter{pict@height}
\newlength{\pict@scale}
\newcommand{\psfigadd}[4]{%
\setcounter{pict@width}{1*\ratio{#2+\pict@scale/2}{\pict@scale}}
\setcounter{pict@height}{1*\ratio{#3+\pict@scale/2}{\pict@scale}}
\setlength{\unitlength}{\pict@scale}
\hbox to #2{\hspace{-\fill}\begin{picture}(\thepict@width,\thepict@height)
\put(0,0){\psfig{figure=#1,width=#2,height=#3,clip=}}
\SetScale{0.283466457}
\SetWidth{1.763889}
{#4}
\end{picture}}
}
\newcounter{pict@widthfst}
\newcounter{pict@widthscd}
\newcounter{pict@widthtot}
\newcommand{\psfigaddtwo}[7]{%
\setcounter{pict@widthfst}{1*\ratio{#2+\pict@scale/2}{\pict@scale}}
\setcounter{pict@widthscd}{1*\ratio{#2+#4+\pict@scale/2}{\pict@scale}}
\setcounter{pict@widthtot}{1*\ratio{#2+#4+#6+\pict@scale/2}{\pict@scale}}
\setcounter{pict@height}{1*\ratio{#3+\pict@scale/2}{\pict@scale}}
\setlength{\unitlength}{\pict@scale}
\hbox{\hspace{-\fill}\begin{picture}(\thepict@widthtot,\thepict@height)
\put(0,0){\psfig{figure=#1,width=#2,height=#3,clip=}}
\put(\thepict@widthscd,0){\psfig{figure=#5,width=#6,height=#3,clip=}}
\SetScale{0.283466457}
\SetWidth{1.763889}
{#7}
\end{picture}}
}
\newcommand{\psfigror}[4]{%
\setcounter{pict@width}{1*\ratio{#2+\pict@scale/2}{\pict@scale}}
\setcounter{pict@height}{1*\ratio{#3+\pict@scale/2}{\pict@scale}}
\setlength{\unitlength}{\pict@scale}
\hbox{\begin{picture}(\thepict@width,\thepict@height)
\put(0,\thepict@height){\psfig{figure=#1,width=#3,height=#2,clip=,angle=270}}
\SetScale{0.283466457}
\SetWidth{1.763889}
{#4}
\end{picture}}
}
\newcommand{\psfigrol}[4]{%
\setcounter{pict@width}{1*\ratio{#2+\pict@scale/2}{\pict@scale}}
\setcounter{pict@height}{1*\ratio{#3+\pict@scale/2}{\pict@scale}}
\setlength{\unitlength}{\pict@scale}
\hbox{\begin{picture}(\thepict@width,\thepict@height)
\put(0,0){\psfig{figure=#1,width=#3,height=#2,clip=,angle=90}}
\SetScale{0.283466457}
\SetWidth{1.763889}
{#4}
\end{picture}}
}
\catcode`\@=12 % @ signs are no longer letters
\newlength\listtextwidth

\catcode`\@=11 % @ signs are now treated as letters
\newlength{\@tabfninsert}
\newlength{\@tabfnwidth}
\newcommand{\tabfootnote}[2]{%
  \setlength{\@tabfninsert}{0.8em}
  \setlength{\@tabfnwidth}{\textwidth}
  \addtolength{\@tabfnwidth}{-\@tabfninsert}
  \addtolength{\@tabfnwidth}{-0.4em}
  \noindent\makebox[\@tabfninsert][r]{\footnotesize$^{#1}$\hfil}\hfill%
  \parbox[t]{\@tabfnwidth}{\footnotesize #2\hfill}}
\catcode`\@=12 % @ signs are no longer letters

\begin{document}
\title{Prompt photons with associated jets in photoproduction at HERA}

%***********************************************************************
% AUTHORS INFORMATION AREA
%***********************************************************************
\author{S.~Chekanov
%
% Optional short acknowledgment: remove next line if non-needed
% \thanks{On behalf of the ZEUS Collaboration.}
%
% DO NOT MODIFY THE FOLLOWING '\vspace' ARGUMENT
\vspace{.3cm}\\
For the ZEUS Collaboration
\vspace{0.2cm}\\
DESY Laboratory, 22607, Hamburg, Germany. \\
On leave from the HEP division, Argonne National Laboratory, \\
9700 S.Cass Avenue,
Argonne, IL 60439, USA \\
E-mail: chekanov@mail.desy.de
}

%
%***********************************************************************
% END OF AUTHORS INFORMATION AREA
%***********************************************************************

\maketitle

\begin{abstract}
Prompt photons, together with an accompanying
jet, have been studied in the photoproduction regime of $ep$ scattering
with the ZEUS detector at HERA.
Predictions based on leading-logarithm
parton-shower Monte Carlo models  and next-to-leading-order (NLO) QCD
underestimate the $\gamma$+jet cross sections for transverse energies
of prompt photons below $7\gev$, while the $k_T$-factorisation QCD calculation
agrees with the data in this region.
\end{abstract}

\section{Theoretical calculations}

Events with an isolated photon (prompt photon) are an important tool
to study hard interaction processes since such photons
emerge without the hadronisation phase.
In particular, final states with a 
prompt photon together with a jet are directly sensitive
to the quark content of the proton through the elastic scattering of a
photon by a quark, $\gamma q\to\gamma q$ (see Fig.~1). 
However, QCD contributions to this lowest-order process lead to significant
sensitivity to the gluon structure function. In particular, 
a contribution to prompt-photon
events from $gq \to q\gamma$ process, in which the photon displays a hadronic
structure (resolved process), is important 
\cite{pr:d52:58,pr:d64:14017,ejp:c21:303}.  Thus,
prompt-photon events
can constrain both proton  
and photon parton densities (PDF).
A number of QCD predictions 
\cite{pr:d52:58, pr:d64:14017, ejp:c21:303, Lipatov:2005tz} 
can be confronted with the data.

\begin{wrapfigure}{r}{0.35\columnwidth}
\centerline{\includegraphics[width=0.25\columnwidth]{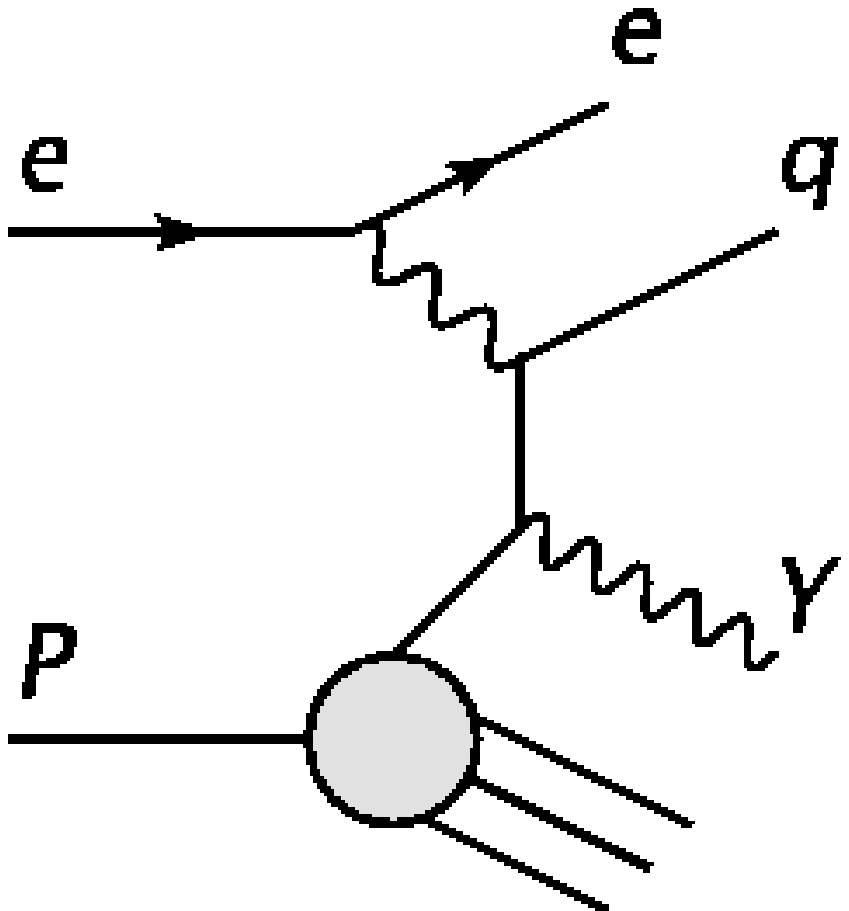}}
Figure~1. Lowest-order diagram (Compton scattering) 
for $\gamma$+jet events in $ep$ collisions.
\end{wrapfigure}

The next-to-leading order (NLO) calculations based on the collinear factorisation  
and the DGLAP formalism
were performed
by Krawczyk and Zembrzuski (KZ) \cite{pr:d64:14017} and 
by Fontanaz, Guillet and Heinrich (FGH)~\cite{ejp:c21:303}. 
No intrinsic transverse momentum of the initial-state
partons in the proton was assumed.
The renormalisation scale for such calculations
was taken to be $\mu_{R}=E_{T}^{\gamma}$,
where $E_{T}^{\gamma}$ is the transverse energy of the photon.
In case of the KZ prediction,  the 
GRV parameterisations  for the proton and  
photon,  as well as for the fragmentation function 
were used~\cite{zfp:c67:433,*pr:d45:3986,*pr:d46:1973,epj:c14:133}.
For the FGH calculation, MRST01 proton PDF 
and the AFG02
photon PDF were used ~\cite{epj:c14:133,*zfp:c64:621}.
The latter calculation
takes into account high-order terms in the QCD expansion which have not been 
considered in the KZ approach.

The QCD calculations based on
the $k_T$-factorisation~\cite{sovjnp:53:657,*np:b366:135,*np:b360:3} approach  
were performed by A.~Lipatov and N.~Zotov (LZ)~\cite{Lipatov:2005tz}.
The  unintegrated quark and gluon
densities of the  proton and photon using the
Kimber-Martin-Ryskin (KMR) prescription \cite{Kimber:2001sc,*Watt:2003mx} were used.
As for the NLO QCD, both the direct and the resolved contributions
were  taken into account.

For all the  calculations discussed above, 
an isolation requirement
$E_{T}^{\gamma}>0.9\, E_{T}^{tot}$ was used, where
$E_{T}^{tot}$ is the total energy of the jet which contains prompt photon. 
Jets were reconstructed with
the longitudinally-invariant
$k_{T}$ algorithm  in inclusive mode \cite{pr:d48:3160,*np:406:187}.
The $\gamma$+jet cross sections were corrected for hadronisation effects
using a Monte Carlo (MC)  simulation.
  
%%%%%%%%%%%%%%%%%%
\section{Event reconstruction}
%%%%%%%%%%%%%%%%%%%%%

Each jet, reconstructed from  energy-flow objects (EFO), was  classified as
either a photon candidate or a hadronic jet.
The photon-candidate jet was required to consist of  EFOs without  associated tracks
and to be within the central tracking detector,
$-0.74<\eta^{\gamma}<1.1$.
For this jet,
$E_{\rm EMC}/E_{\rm tot}>0.9$ is  required, where  $E_{\rm EMC}$ is
the energy reconstructed in the electromagnetic part of the CAL and
$E_{\rm tot}$ is the total energy of this jet.
After correction for energy losses, the cut $E_{T}^{\gamma}>5\gev$
was applied.

\begin{center}
\vspace{-0.5cm}
\begin{minipage}[c]{0.48\textwidth}
\includegraphics[width=6.8cm,angle=0]{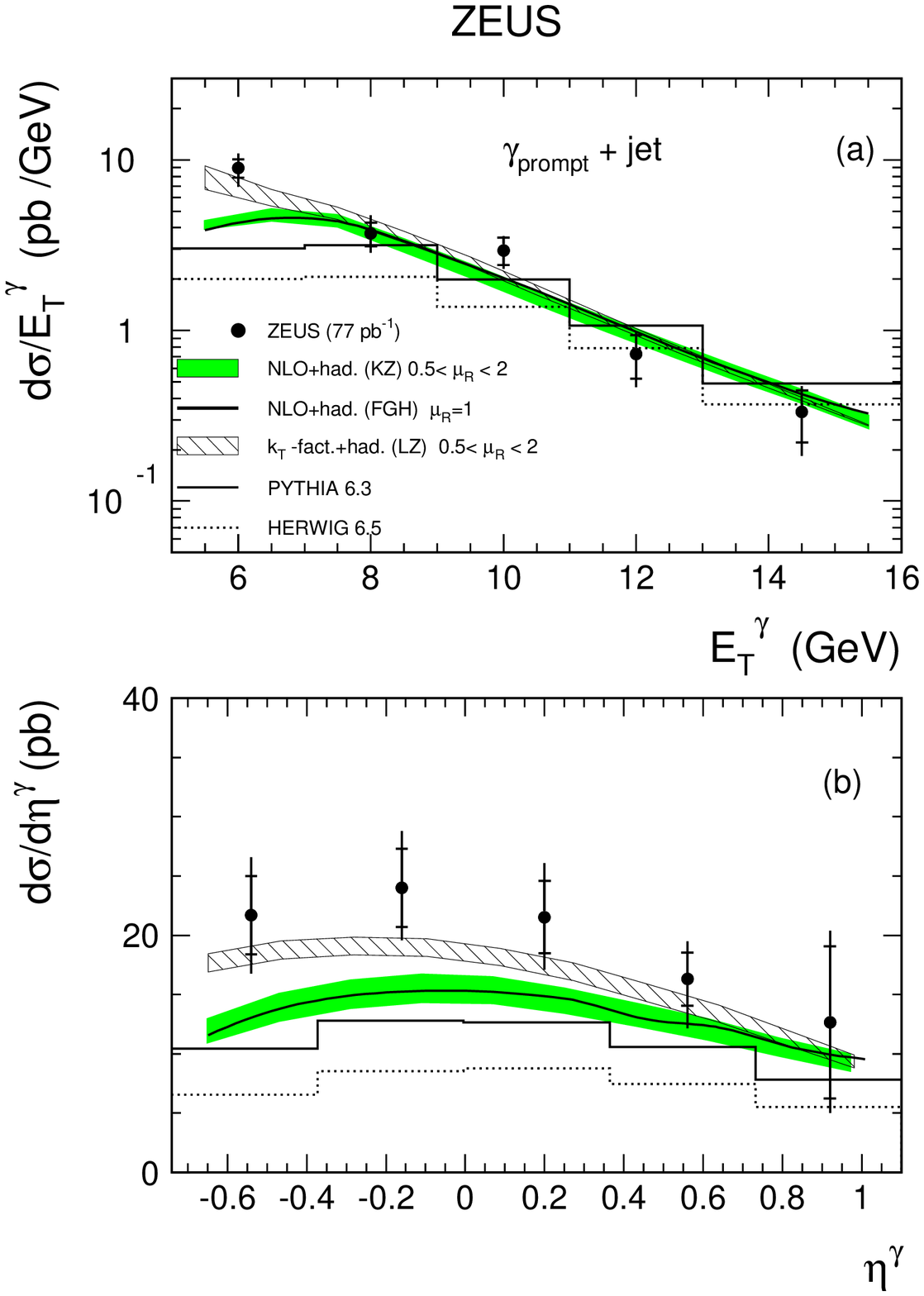}
\label{theta1}
\end{minipage}
\hfill
\begin{minipage}[c]{.48\textwidth}
\includegraphics[width=6.8cm,angle=0]{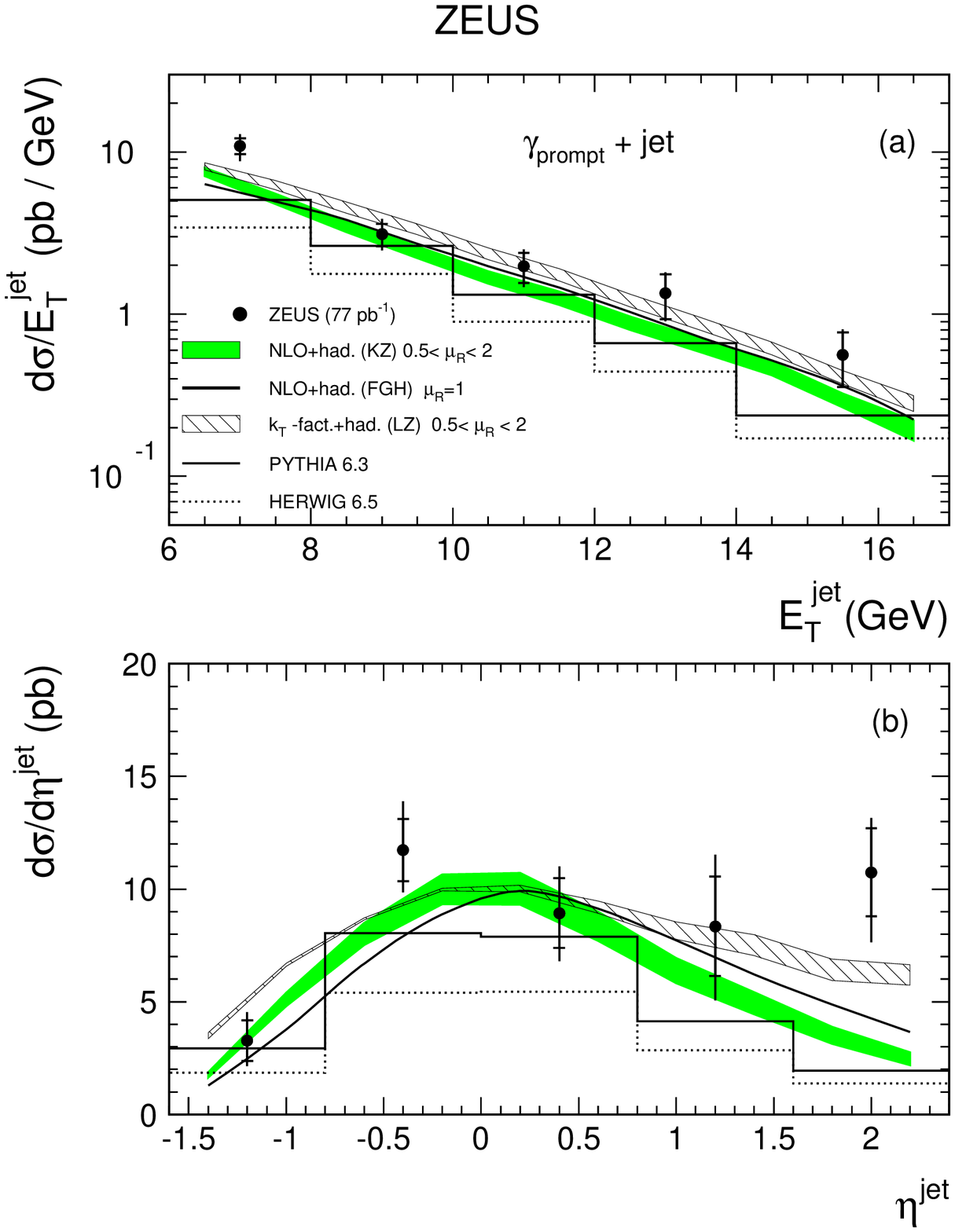}
\label{theta2}
\end{minipage}
\end{center}
\vspace{-0.5cm}
Figure~1. The differential $\gamma$+jet cross sections 
as functions of $E_{T}$ and $\eta$  of  the prompt
photon and the jet, as described in the figure.
The data are compared to QCD  calculations
and MC models.
The shaded bands correspond to a typical  renormalisation
scale uncertainty which was obtained by changing $\mu_R$ by a factor of 0.5 and 2.
\vspace{0.5cm}

\begin{center}
\vspace{-1.0cm}
\begin{minipage}[c]{0.48\textwidth}
\includegraphics[width=6.1cm,angle=0]{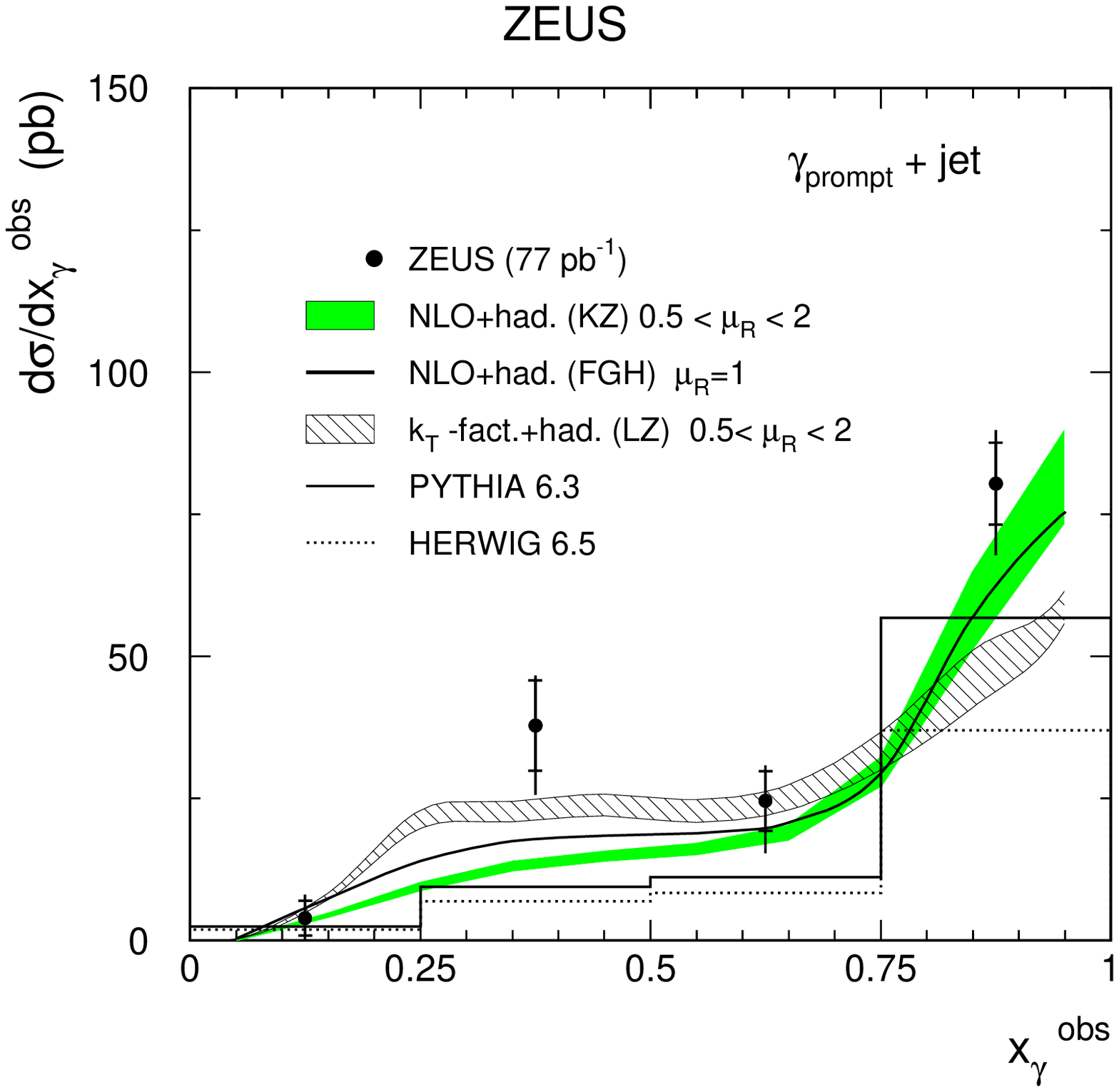}
\label{theta1a}
\end{minipage}
\hfill
\begin{minipage}[c]{.48\textwidth}
\includegraphics[width=6.1cm,angle=0]{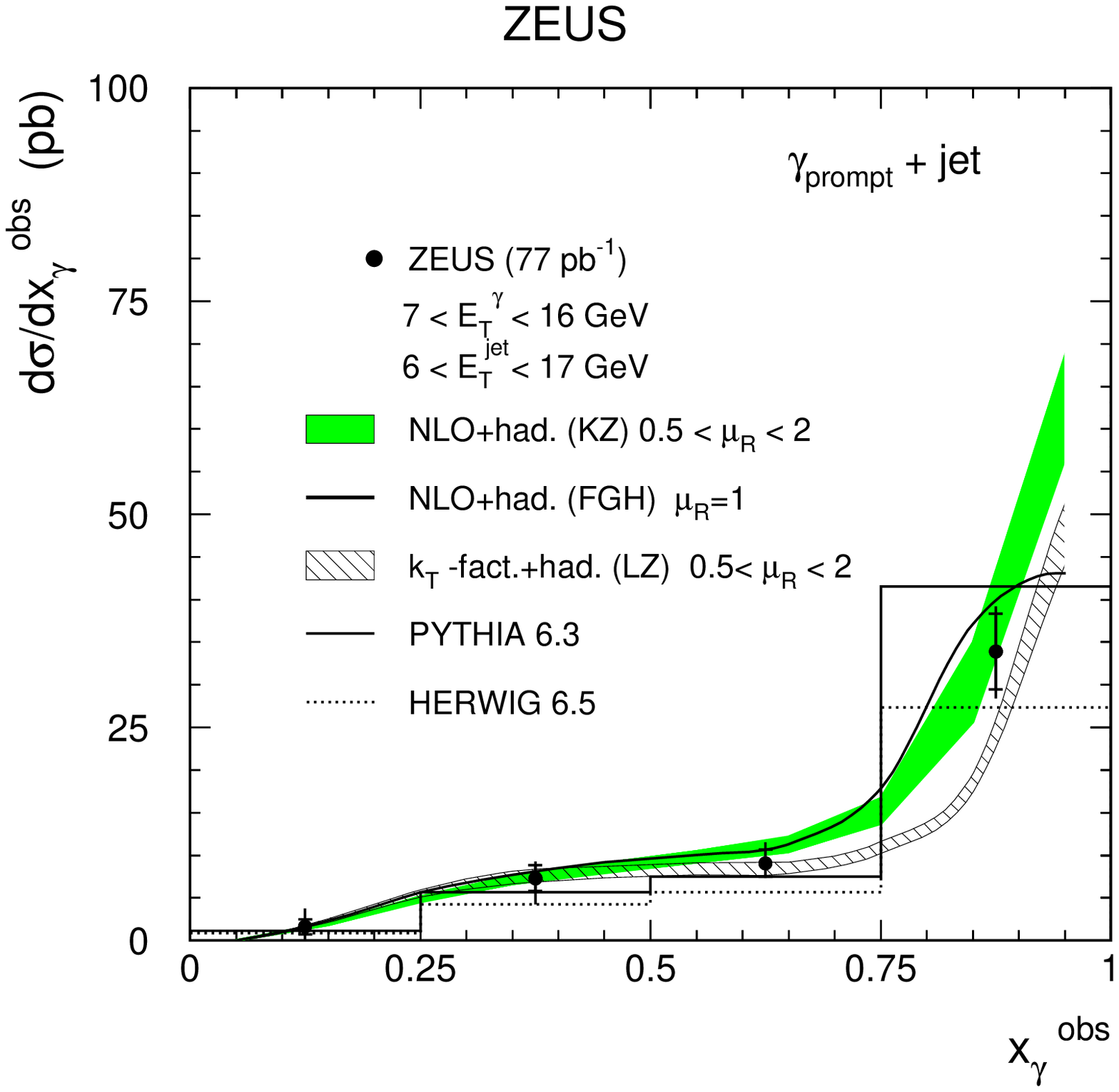}
\label{theta2a}
\end{minipage}
\end{center}
\vspace{-0.5cm}
Figure~2. The  $x_{\gamma}^{\mathrm{obs}}$  cross section for $\gamma$+jet events
compared to the NLO QCD calculations
and MC  models for $E_{T}^{\gamma}>5\gev$ (left) and $E_{T}^{\gamma}>7\gev$ (right). 
\vspace{0.5cm}

Hadronic jets, after correction for energy losses,
were  selected in the kinematic range $E_{T}^{\rm jet}>6\gev$,
$-1.6<\eta^{\rm jet}<2.4$. 
If more than one jet was found within the above kinematic cuts,
the jet with the highest $E_{T}^{\rm jet}$ was selected.

For the prompt-photon identification,
the conversion-probability method was used~\cite{Chekanov:2006un}.
In contrast to the shower-profile approach adopted in previous HERA measurements, 
the present
approach uses the probability
of conversion of photons to  $e^{+}e^{-}$ pairs in
detector elements and inactive material (mainly the ZEUS superconducting coil) 
in front of the barrel calorimeter (BCAL).
Since the conversion probability for a single photon is
smaller than for multiphoton events arising from 
neutral meson decays ($\pi^0$, $\eta$, etc.),
one can extract the $\gamma$ signal by performing a statistical
background subtraction.  

To determine the number of charged particles in the photon shower,
the ZEUS barrel preshower detector (BPRE) \cite{magill:bpre}
located in front of the BCAL was used.
The measured output, calibrated in units of 
minimum ionising particle (mips), is proportional
to the energy loss of the incident particle after interaction with inactive material.
The response of the BPRE to single isolated photons was verified 
using deeply virtual
Compton scattering events. 
For the $\gamma$+jet, the  BPRE signal for the $\gamma$ candidates 
was fitted using a MC  model
with and without prompt photons, and the number of events
associated with the photon signal was extracted.

%%%%%%%%%%%%%%%%%%%%%%%%%%%%%
\section{Results and conclusions}

The total cross section for the process $ep\to e+\gamma_{\rm prompt}+\mathrm{jet}+X$
for $0.2<y<0.8$, $Q^{2}<1\,\mathrm{GeV^{2}}$, $5<E_{T}^{\gamma}<16$
GeV, $6<E_{T}^{\rm jet}<17$ GeV, $-0.74<\eta^{\gamma}<1.1$,  $-1.6<\eta^{\rm jet}<2.4$
and  $E_{T}^{\gamma, \rm (true)}>0.9\, E_{T}^{\gamma}$  was measured to be
$
\sigma(ep\to e+\gamma_{\rm prompt}+\mathrm{\rm jet}+X)=33.1\pm 3.0\,(\mathrm{stat.})\,_{-4.2}^{+4.6}(\mathrm{syst.})
\:\mathrm{pb.}
$

This value agrees well with the LZ calculation
($30.7^{+3.2}_{-2.7}\pb$), but is higher than for the NLO QCD 
($23.3^{+1.9}_{-1.7}\pb$ (KZ)
and $23.5^{+1.7}_{-1.6}\pb$ (FGH)) and MC models. 
 
The differential cross sections as functions of $E_{T}$ and $\eta$ for the prompt-photon
candidates and for the accompanying jets are shown in Figure~1.
The MC  differential cross sections do not rise as steeply
at low $E_{T}^{\gamma}$ as do the data. 
The  KZ NLO prediction  
describes the data better.
However,  it underestimates the observed cross section at low
$E_{T}^{\gamma}$ and in the forward jet region.
The FGH prediction is similar to the KZ NLO.
The LZ prediction based on the $k_T$-factorisation approach
gives the best description
of the $E_{T}$ and $\eta$ cross sections.

Figure~2(left)  shows the  distribution for
$x_{\gamma}^{\rm obs}$  defined as $\sum_{\gamma, \rm jet} (E_i-P_Z^i)/(2E_e y)$
(the sum runs over the photon candidate and the hadronic jet).
The difference between the NLO QCD and the data is mainly concentrated in the
resolved photon region.

It is important 
to verify the level of agreement with NLO when the minimum transverse energy of the detected
prompt photons is increased from $5\gev$ to $7\gev$.
In comparison with previous measurements,
such a choice may emphasize different aspect of contributions
of high-order QCD radiation,
since the transverse energy of the prompt-photon is larger
than that of the jet.

Figure~2(right) shows the corresponding
$x_{\gamma}^{\rm obs}$ distribution.  
For the $E_{T}^{\gamma}>7\gev$ cut, both the NLO QCD and the LZ predictions
agree well with the data. 
There is also good agreement for the $E_T$ and $\eta$ kinematic variables~\cite{Chekanov:2006un}. 
 
{\it Acknowledgements.}
I thank M.~Fontannaz, G.~Heinrich, M.~Krawczyk, A.~Lipatov, N.~Zotov 
and A.~Zembrzuski for discussions and
for providing the QCD calculations.

% ****************************************************************************
% BIBLIOGRAPHY AREA
% ****************************************************************************
%%%%%%%%%%%%%%%%%%%%%% references %%%%%%%%%%%%%%%%%%%%%%%%%%%%%%
% draft type, with title and hp-ex
%%%%%%%%%%%%%%%%%%%%%% references %%%%%%%%%%%%%%%%%%%%%%%%%%%%%%
\bibliographystyle{./l4z_default}
\def\bibname{\Large\bf References}
\def\refname{\Large\bf References}
\pagestyle{plain}
\bibliography{chekanov_sergei_prph}

\providecommand{\etal}{et al.\xspace}
\providecommand{\coll}{Coll.\xspace}
\catcode`\@=11
\def\@bibitem#1{%
\ifmc@bstsupport
  \mc@iftail{#1}%
    {;\newline\ignorespaces}%
    {\ifmc@first\else.\fi\orig@bibitem{#1}}
  \mc@firstfalse
\else
  \mc@iftail{#1}%
    {\ignorespaces}%
    {\orig@bibitem{#1}}%
\fi}%
\catcode`\@=12
\begin{mcbibliography}{10}

\bibitem{pr:d52:58}
L.~Gordon, W.~Vogelsang,
\newblock Phys.\ Rev.{} {\bf D52},~58~(1995)\relax
\relax
\bibitem{pr:d64:14017}
M.~Krawczyk, A.~Zembrzuski,
\newblock Phys.~Rev.{} {\bf D64},~14017~(2001)\relax
\relax
\bibitem{ejp:c21:303}
M.~Fontannaz, J.~P. Guillet, G.~Heinrich,
\newblock Eur.~Phys.~J.{} {\bf C21},~303~(2001)\relax
\relax
\bibitem{Lipatov:2005tz}
A.~Lipatov, N.~Zotov,
\newblock Phys.~Rev.{} {\bf D72},~054002~(2005)\relax
\relax
\bibitem{zfp:c67:433}
M.~Gl\"uck, E.~Reya, A.~Vogt,
\newblock Z.\ Phys.{} {\bf C67},~433~(1995)\relax
\relax
\bibitem{pr:d45:3986}
M.~Gl\"uck, E.~Reya, A.~Vogt,
\newblock Phys.\ Rev.{} {\bf D45},~3986~(1992)\relax
\relax
\bibitem{epj:c14:133}
A.~D. Martin, et~al.,
\newblock Eur.\ Phys.\ J.{} {\bf C14},~133~(2000)\relax
\relax
\bibitem{zfp:c64:621}
P.~Aurenche, J.~P. Guillet, M.~Fontannaz,
\newblock Z.\ Phys.{} {\bf C64},~621~(1994)\relax
\relax
\bibitem{sovjnp:53:657}
E.~M. Levin, et~al.,
\newblock Sov.\ J.\ Nucl.\ Phys.{} {\bf 53},~657~(1991)\relax
\relax
\bibitem{np:b366:135}
S.~Catani, M.~Ciafaloni, F.~Hautmann,
\newblock Nucl.\ Phys.{} {\bf B366},~135~(1991)\relax
\relax
\bibitem{np:b360:3}
J.~Collins, R.~Ellis,
\newblock Nucl.\ Phys.{} {\bf B360},~3~(1991)\relax
\relax
\bibitem{Kimber:2001sc}
M.~A. Kimber, A.~D. Martin, M.~G. Ryskin,
\newblock Phys. Rev.{} {\bf D63},~114027~(2001)\relax
\relax
\bibitem{Watt:2003mx}
G.~Watt, A.~D. Martin, M.~G. Ryskin,
\newblock Eur. Phys. J.{} {\bf C31},~73~(2003)\relax
\relax
\bibitem{pr:d48:3160}
S.~Ellis, D.~Soper,
\newblock Phys.~Rev.{} {\bf D48},~3160~(1993)\relax
\relax
\bibitem{np:406:187}
S.~Catani, et~al.,
\newblock Nucl.\ Phys.{} {\bf B406},~187~(1993)\relax
\relax
\bibitem{Chekanov:2006un}
ZEUS~Collaboration, S.~Chekanov, et~al.,
\newblock Eur. Phys. J.{} {\bf C49},~511~(2007)\relax
\relax
\bibitem{magill:bpre}
S.~Magill, S.~Chekanov,
\newblock {\em Proceedings of the IX Int. Conference on Calorimetry (Annecy,
  Oct 9-14, 2000)}, B.~Aubert, et~al.~(eds.), p.~625.
\newblock Frascati Physics Series 21, Annecy, France (2001)\relax
\relax
\end{mcbibliography}

\end{document}